\documentclass[preprint,showpacs,amsmath,amssymb]{revtex4}
\usepackage{mathrsfs}

\usepackage{graphicx,color}
\usepackage{dcolumn}
\usepackage{bm}

\newcommand{\nb}{\nonumber}
\usepackage{graphicx}% Include figure files
\usepackage{dcolumn}% Align table columns on decimal point
\usepackage{bm}% bold math
\def\be{\begin{equation}}
\def\ee{\end{equation}}
\def\bea{\begin{eqnarray}}
\def\eea{\end{eqnarray}}

\begin{document}

%-------------------------------------------------------------------------------------------------------

\title{Next-to-leading order QCD corrections to $\chi_{cJ} W^+ b $ associated production from top-quark decay}
\author{Zhou Chao$^1$}%\email{zhouchao@ahu.edu.cn}
\author{Li Gang$^1$}\email{lig2008@mail.ustc.edu.cn}
\author{Song Mao$^1$}%\email{songmao@ahu.edu.cn}
\author{Ma Wen-Gan$^2$}%\email{mawg@ustc.edu.cn}
\author{Zhang Ren-You$^2$}%\email{zhangry@ustc.edu.cn}

\affiliation{$^1$ School of Physics and Materials Science, Anhui University, Hefei, Anhui 230039,People's Republic of China}
\affiliation{$^2$ Department of Modern Physics, University of Science and Technology of China (USTC),Hefei, Anhui 230026, People's Republic of China}
\date{\today }

%-------------------------------------------------------------------------------------------------------
\begin{abstract}
We calculate the next-to-leading order QCD corrections to the excited charmonium $\chi_{cJ}$ associated
with $ W^+ b $ production from top-quark decay. Our results show that detecting the
$\chi_{c0}$ production from top-quark decay is very
difficult, but the $\chi_{c1}$ and $\chi_{c2}$ productions have the potential to be
detected at the LHC. If the prompt $\chi_{cJ}$ production from top-quark decay is really
detected at the LHC, it will be useful not only for investigating $J/\psi$ production from
top-quark decay but also for understanding the heavy quarkonium production mechanism.

\end{abstract}

\pacs{12.38.Bx, 14.40.Pq, 14.65.Ha} \maketitle

\section{Introduction}
\par
Heavy quarkonium is a multiscale system, which offers a good testing ground for
investigating the QCD in both perturbative and nonperturbative regimes.
The factorization formalism of nonrelativistic QCD
(NRQCD) \cite{bbl} as a rigorous theoretical framework to describe the
heavy quarkonium production and decay has been widely investigated both in
experimental and theoretical aspects. People believe that NRQCD, the only effective
field theory allowing for consistent QCD-based calculations beyond the Born
approximation, may be the most promising theory to describe the heavy quarkonium physics.

Through efforts from both the experimental and theoretical sides, substantial
progress has been achieved in heavy quarkonium physics, many processes have been
calculated to next-to-leading order(NLO) in $\alpha_s$
\cite{longma,longbu,longgong,Butenschoen:2009zy,Butenschoen:2011ks,Ma:2010vd,
Butenschoen:2012px,Chao:2012iv,Shao:2014fca,Shao:2014yta,Klasen:2004tz,Klasen:2004az,
Butenschoen:2011yh}. For the prompt $J/\psi$ production, almost all the relevant
observable predictions are available at the NLO, and based on different philosophies,
the color-octet(CO) long-distance
matrix elements (LDMEs) of $J/\psi$ have been extracted by three groups
independently. It has been found that for quarkonium
production and decay in the framework of NRQCD in many cases the leading order(LO)
calculation is inadequate and the NLO QCD corrections are crucial.
The discrepancies between LO calculations and unpolarized experimental results are
fairly well described by the NRQCD theory through including higher order
corrections\cite{d1,d2,comtev,Chao:2012iv,pnloz}. But for polarization production,
though the NLO corrections have been considered, people are not able to fully explain
the polarization production by theoretical analyses in a way consistent with the
world data on the unpolarized yield, and the polarization puzzle still poses a
challenge to the heavy quarkonium physics\cite{polar1,heRE}.

Therefore,a further test the mechanism of quarkonium production is needed, and more
processes of heavy quarkonium production and decay should be investigated. The
study of the production for excited charmonium other than $J/\psi$ may also be valuable,
not only is the study of excited heavy quarkonium production important for $J/\psi$
production for the excited heavy quarkonium to be able to radiatively decay to $J/\psi$, but also the
study of excited heavy quarkonium production itself can directly deepen our understanding about
QCD. Many processes of $\chi_{cJ}$ $(J=0,1,2)$ production at the Tevatron
and LHC have been studied up to the NLO, and the
$r=\frac{m^2_c<{\cal O}^{\chi_{c0}}[^3S_1^{(8)}]>}{<{\cal O}^{\chi_{c0}}[^3P_0^{(1)}]>} $
has been given by using the Tevatron data, LHCb data and CMS data. But its accuracy
is not very satisfying; the $r$ value can be varied from $0.21$ to $0.35$ in fitting different
data with different hypotheses \cite{Ma:2010vd,Shao:2014fca}.

\par
At the LHC, The latest  estimations for
$\sigma(pp \to t\bar{t}X)$ range from  $874^{+14}_{-33}$ pb~\cite{Langenfeld:2009tc} to
$943 \pm 4({\rm kinematics}) ^{+77}_{-49}({\rm scale})\pm 12 ({\rm PDF})$ pb~\cite{Kidonakis:2008mu}
for $m_t=173$ GeV and $\sqrt{s}=14$ TeV. Therefore, it is significant to perform detailed study
of heavy quarkonium production from top-quark decay at the LHC. Many heavy quarkonium productions from
top-quark decay processes have been calculated at the LO\cite{Liao:2013ika,Qiao:1996rd,qiao,Yuan:1997yh},
and the decay widths of top quark to S-wave $\bar{b}c$, $\bar{c}c$ and $\bar{b}b$ bound states at the NLO
are available now\cite{Sun:2010rw,litop}. For the investigation of
$t\to\chi_{cJ}+ W^+ +b $ processes may provide important information not only for the $J/\psi$ associated
with $ W^+ b $ production from top-quark decay but also can provide an excellent platform to extract
the universality LDMEs of $\chi_{cJ}$. In this paper, we will calculate the $t\to\chi_{cJ}+ W^+ +b $
processes up to the NLO in $\alpha_s$ within the NRQCD framework by applying the covariant projection method\cite{p2}.
The paper is organized as follows. in Sec.II, we present the details of the calculation strategies, and Sec.III
is arranged to present the numerical results. Finally, a short summary and discussions
are given.

\section{Calculation descriptions}
\par
In this section, we present the calculation about the decay width for processes
$t\to\chi_{cJ}+ W^+ +b+X$ to the NLO of $\alpha_s$. At the LO, only the $^3S_1^{(8)}$
Fock state has made a contribution, and the Feynman diagrams for this partonic
process are drawn in Figs. 1(a) and 1(b). In the nonrelativistic limit, the short-distance
coefficients of $^3S_1^{(8)}$ Fock state for processes $t\to\chi_{cJ}+ W^+ +b+X$ are
the same as the process $t\to J/\psi+ W^+ +b+X$ in $^3S_1^{(8)}$ Fock state at the LO.
Applying the covariant-projector method of Ref.\cite{p2}, we can get analytic
short-distance coefficients of processes $t\to\chi_{cJ}+ W^+ +b+X$, which are the
same as in Refs.\cite{qiao,litop}.

\begin{figure}[!htb]
\begin{center}
\begin{tabular}{cc}
{\includegraphics[width=0.6\textwidth]{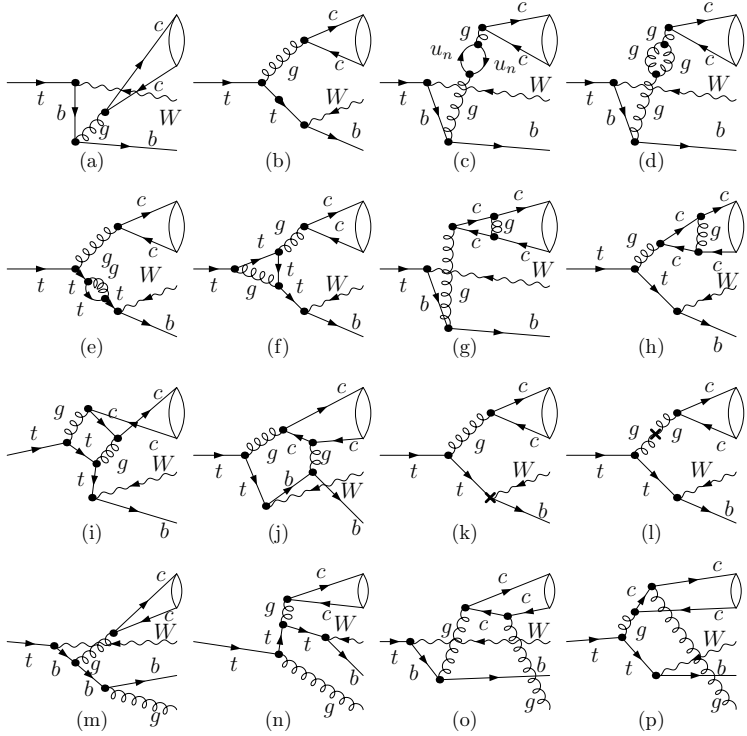}}
\end{tabular}
\end{center}
 \vspace*{-0.7cm}\caption{Some representative LO and QCD NLO
Feynman diagrams for the $t\to\chi_{cJ}+ W^+ +b$ decay processes.Where
(a) and (b) are tree level diagrams, (c)-(j) are part loop diagrams,
(k)-(l) are part counterterms diagrams and (m)-(p) are part real
gluon emission diagrams for $t\to\chi_{cJ}+ W^+ +b$ decay process.} \label{f1}
\end{figure}

\par
The NLO QCD corrections contain the virtual correction and the real
gluon emission correction; the former is only related to $^3S_1^{(8)}$
Fock-state contribution, and the latter involves $^3S_1^{(8)}$ and
$^3P_J^{(1)}$ Fock-state contributions. Some representative loop and real gluon emission
Feynman diagrams for the $t\to\chi_{cJ}+ W^+ +b+X$ decay processes at
the NLO are presented in Figs. 1(c)-1(p). There exist UV, soft and coulomb
singularities in virtual correction, and soft singularities will emerge
from both the $^3S_1^{(8)}$ Fock-state and $^3P_J^{(1)}$ Fock-state
contributions when we calculate the real gluon emission process. The UV
divergences from the virtual correction are removed after
renormalization procedure. The soft divergences from the
one-loop diagrams will be canceled by similar singularities from
the $^3S_1^{(8)}$ Fock-state contribution of soft real gluon emission.
Nevertheless, it still contains coulomb
singularities in virtual correction and soft singularities
arising from the $^3P_J^{(1)}$ Fock-state contribution of real gluon
process. These singularities are not infrared divergence in
the usual sense, and they can only be eliminated in the spirit of
the factorization approach, by taking the corresponding
corrections to the operator $<{\cal O}^{\chi_{cJ}}[^3S_1^{(8)}]>$
into account. In Fig. 2, we present the divergence
structure and divergence cancellation routes in the NLO
calculation for the $t\to\chi_{cJ}+ W^+ +b$.

\begin{figure}[!htb]
 \vspace*{-0.3cm}
\begin{center}
\begin{tabular}{cc}
{\includegraphics[bb = 10 115 155 190,scale = 1.5]{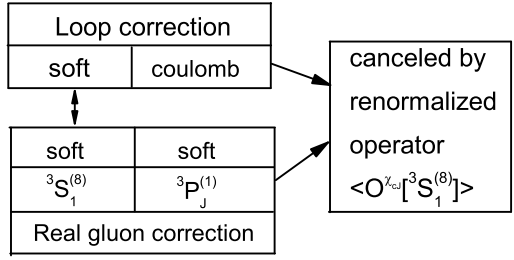}}
\end{tabular}
\end{center}
 \vspace*{-0.5cm}\caption{The IR and Coulomb
singularity structures in the NLO QCD calculations for the $t\to\chi_{cJ}+ W^+ +b$ decay processes.} \label{f2}
\end{figure}

\par
In our calculations, the dimensional regularization scheme is adopted to regularize the UV and IR divergences under t'Hooft-Feynman gauge. We use the modified
minimal subtraction ($\overline{{\rm MS}}$) and on-mass-shell schemes to
renormalize the strong coupling constant and the quark wave functions, respectively.
A small relative velocity $v$ between $c$ and $\bar{c}$ has been used to regularize
coulomb singularities\cite{coulomb} for Fig. 1(g) and 1(h) in the virtual correction calculation. Meanwhile, the phase space slicing (PSS) method
\cite{pss} has been employed for dealing with the soft singularities in real gluon
emission corrections.

\par
When we calculate the short-distance coefficients of $^3S_1^{(8)}$ Fock state in the
virtual and real gluon emission contribution, we use the strategy same as in Ref.\cite{litop}.
The detail for the calculations of $^3S_1^{(8)}$ Fock-state contributions is too tedious to
be presented here for the sake of brevity. As for the $^3P_J^{(1)}$ Fock-state decay contributions
to the real gluon emission processes $t\to \chi_{cJ}+ W^+ +b+ g$, we denote the decay processes as
\begin{equation}\label{emission}
t(p_1) \to \chi_{cJ}(p_2) +W^+ (p_3) + b(p_4) +  g(p_5),
\end{equation}
and some of the Feynman diagrams for these partonic processes are presented in
Fig. 1(o) and 1(p).

There are soft singularities arising from the real gluon processes with
$^3P_J^{(1)}$ Fock state, which can be
isolated by slicing the phase space into two different
regions based on PSS method. By introducing a small cutoff $\delta_s$, the phase
space of $t\to \chi_{cJ}+ W^+ +b+ g$ is separated into two regions, according to whether
the emitted gluon is soft, i.e., $E_5 < \delta_s m_{t}/2$,
or hard, i.e., $E_5 \ge \delta_sm_{t}/2$. In our numerical calculations, when $\delta_s$ varies from
$10^{-5}$ to about $10^{-4}$, the $\Delta \Gamma_{soft_{^3P_J^{(1)}}}$ and $\Delta
\Gamma_{hard_{^3P_J^{(1)}}}$ remain almost unchanged at the value of about 1.0\% of the total width.
In further calculations, we set $\delta_s=1\times
10^{-4}$. Then the decay width for the processes $t\to \chi_{cJ}+ W^+ +b+ g$
with the $^3P_J^{(1)}$ Fock state can be expressed as
\begin{eqnarray}
\Delta \Gamma_{real_{^3P_J^{(1)}}}(t\to \chi_{cJ}+ W^+ +b+ g) &=&
\Delta \Gamma_{soft_{^3P_J^{(1)}}}(t\to \chi_{cJ}+ W^+ +b+ g)\nb\\
&+&\Delta \Gamma_{hard_{^3P_J^{(1)}}}(t\to \chi_{cJ}+ W^+ +b+ g).
\end{eqnarray}
$\Delta \Gamma_{hard_{^3P_J^{(1)}}}$ is finite and
can be integrated in four dimensions by using the Monte Carlo method.
Using the method of Ref.\cite{p2}, we can get the expression
of $\Delta\Gamma_{soft_{^3P_J^{(1)}}}$ as
\begin{eqnarray}
\label{collinear-d}
\Delta \Gamma_{soft_{^3P_J^{(1)}}}&=&-\left(\frac{1}{\epsilon}-2{\rm
ln}\delta_s+\frac{1}{\beta}{\rm
ln}\frac{1+\beta}{1-\beta}\right)\frac{4 \alpha_s C_F}{3\pi
m^2_c}\nb \\
&\times&\frac{\Gamma(1-\epsilon)}{\Gamma(1-2 \epsilon)}\left(\frac{4
\pi
\mu_r^2}{\hat{s}}\right)^{\epsilon} <{\cal O}^{\chi_{cJ}}\left[{}^3\!P_J^{(1)}\right]> \nb\\
&\times&\frac{\Gamma_{LO}}{<{\cal
O}^{\chi_{cJ}}[^3S_1^{(8)}]>} \nb\\
\end{eqnarray}
with $\beta=\sqrt{1-4 m_c^2/E_2^2}$ and where $E_2$ is the energy of $\chi_{cJ}$ and $C_F=\frac{N_c^2-1}{2N_c}=\frac{4}{3}$,
for quark colors $N_c=3$.

When we deal with the renormalization of the color octet
$^3S_1^{(8)}$ LDME, we adopt the same method as in Ref.~\cite{Klasen:2004tz}.
\begin{eqnarray}
<{\cal O}^{\chi_{cJ}}[^3S_1^{(8)}]>_{Born}
&=&<{\cal O}^{\chi_{cJ}}[^3S_1^{(8)}]>_r(\mu_{\Lambda})
\left[1-\left(C_F-\frac{C_A}{2}\right)\,\frac{\pi\alpha_s}{2v}\right]
\nonumber\\
&&+\frac{4\alpha_s}{3\pi m_c^2}
\left(\frac{4\pi\mu_r^2}{{\mu_{\Lambda}}^2}\right)^\epsilon\exp(-\epsilon\gamma_E)
\frac{1}{\epsilon}\nonumber\\
&&{}\times\sum_{J=0}^2\left(
C_F<{\cal O}^{\chi_{cJ}}[^3P_J^{(1)}]>
+B_F<{\cal O}^{\chi_{cJ}}[^3P_J^{(8)}]>\right),
\label{eq:ren}
\end{eqnarray}

where $\mu_r$ is the t'Hooft mass scale and $B_F=\frac{N_c^2-4}{4N_c}=\frac{5}{12}$.
$\mu_{\Lambda}$ is the NRQCD scale, where $v=|\overrightarrow{p_{c}}-\overrightarrow{p_{\bar c}}|/m_c$,
defined in the meson rest frame, and we use this small relative velocity between $c$ and $\bar{c}$ to
regularize the Coulomb singularities. Using the strategy shown in Fig. 2,
after taking into account the NRQCD NLO corrections to the operator $<{\cal O}^{\chi_{cJ}}[^3S_1^{(8)}]>$,
all the IR and Coulomb singularities can be cancelled,  and we can get the finite NLO QCD corrected
total decay width for the processes $t\to \chi_{cJ}+ W^+ +b$. Then, the $t\to \chi_{cJ}+ W^+ +b$ total
decay width including the NLO QCD corrections can be obtained by summing all the contribution
parts:
\begin{eqnarray}
\Gamma_{NLO} &=& \Gamma({^3S_1^{(8)}})+\Gamma(^3P_J^{(1)}) \nb\\
             &=& \Gamma_{LO}({^3S_1^{(8)}})+\Delta \Gamma_{Virtual}({^3S_1^{(8)}}) +\Delta \Gamma_{Real}(^3S_1^{(8)})  +\Delta \Gamma_{Real}(^3P_J^{(1)})
\end{eqnarray}

\section{Numerical results and discussion}
\par
In the numerical calculations, we
use one-loop and two-loop running $\alpha_s$  in the LO and NLO
calculations, respectively, which means $\alpha_s(M_Z) = 0.130$ and $\alpha_s(M_Z) = 0.118$ for the LO and NLO
calculations, respectively. The relevant quark masses and fine structure constant are
taken as: $m_q=0\ (q=u,d,s)$, $m_c=m_{\chi_{cJ}}/2=1.5~GeV$,
$m_W=80.398~GeV$, $m_b=4.75~GeV$, $m_t=173~GeV$ and $\alpha
=1/137.036$. The renormalization and NRQCD scales are chosen as
$\mu_r=m_t$ and $\mu_{\Lambda}=m_c$, respectively.

Following the heavy-quark spin symmetry, the multiplicity
relations of LDMEs
\begin{eqnarray}
<{\cal O}^{\chi_{cJ}}[{}^3\!P_J^{(1)}]> &=&(2J+1) <{\cal
O}^{\chi_{c0}}[{}^3\!P_0^{(1)}]>,
\nonumber\\
<{\cal O}^{\chi_{cJ}}[{}^3\!S_1^{(8)}]> &=&(2J+1) <{\cal
O}^{\chi_{c0}}[{}^3\!S_1^{(8)}]> \label{eq:mul}
\end{eqnarray}
can be assumed satisfied \cite{bbl}. The relation between the color-singlet (CS) matrix elements $<
O^{\chi_{c0}}[{}^3P_0^{(1)}]> $
 of $\chi_{c0}$ and the P-wave function at the origin can be written as the
 formula $< O^{\chi_{c0}}[{}^3P_0^{(1)}]>
=\frac{3N_c}{2\pi}|R_P^{\prime}(0)|^2$. In predicting the production cross sections,
$|R_P^{\prime}(0)|^2=0.075 GeV^{5}$ from the potential model
calculations \cite{Eichten:1995PRD} has been used in our calculation.

\begin{figure}[!htb]
\begin{center}
\begin{tabular}{cc}
{\includegraphics[width=0.6\textwidth]{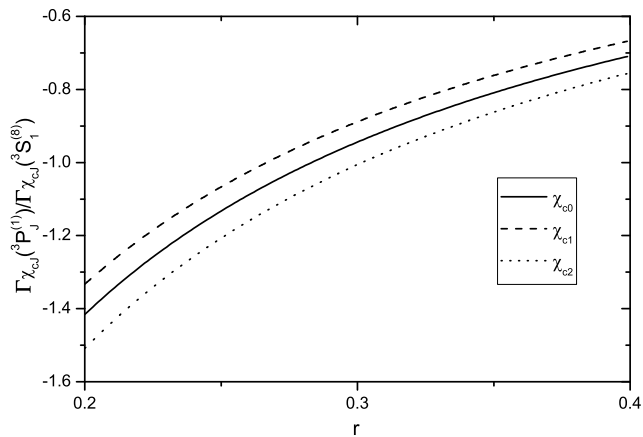}}
\end{tabular}
\end{center}
 \vspace*{-0.7cm}\caption{The ratio of $\Gamma_{^3P_J^{(1)}}$  to $\Gamma_{^3S_1^{(8)}}$ at
 NLO as a function of the the matrix element ratio $r$ for the processes $t\to \chi_{cJ}+ W^+ +b$.} \label{f3}
\end{figure}

 As for the CO matrix element $<O^{\chi_{c0}}[{}^3S_1^{(8)}]> $, though the NLO
results have been obtained based on fitting the Tevatron data, LHCb data, and CMS data,
its accuracy is very poor. The ratio
$r=\frac{m^2_c<{\cal O}^{\chi_{c0}}[^3S_1^{(8)}]>}{<{\cal O}^{\chi_{c0}}[^3P_0^{(1)}]>} $
is acceptable with $r=0.27\pm 0.06$ when fitting Tevatron data and CMS data. However, for the LHCb
data, the value of $r$ can be varied from $0.35$ to $0.31$ when using a different $p_T$ cutoff. In
Fig. 3, we present
the ratio of $\Gamma_{^3P_J^{(1)}}$  to $\Gamma_{^3S_1^{(8)}}$ at the NLO as a function of
the the matrix element ratio $r$ for the processes $t\to \chi_{cJ}+ W^+ +b$. In our
calculations, we have not fixed the value of $r$ and varied it from 0.20 to 0.40. From
Fig. 3, we can see that the contribution from $^3P_J^{(1)}$ Fock state is very important
and surprising with negative sign. This negative
contribution of $^3P_J^{(1)}$ state is mainly due to the fact we have used the $\overline{MS}$ subtraction scheme\cite{Ma:2010vd} in renormalizing the NRQCD LDMEs $<{\cal
O}^{\chi_{cJ}}[^3S_1^{(8)}]>$ and set factorization scale $\mu_\Lambda=m_c$. The $^3P_J^{(1)}$
Fock state contributions mainly come from the real gluon contribution
and the operator contribution induced by the mixing
of the $c\bar{c}$ Fock state NRQCD operators at one loop level
[12], where the latter depends on the subtraction scheme
and the factorization scale $\mu_\Lambda$ in the NLO calculation.
But this dependence will be compensated by the corresponding
one of the $^3S_1^{(8)}$ state contribution in the calculation of the total decay ratio $\Gamma_{\chi_{cJ}}$. From Eq(4), we can see that if we select
a smaller $\mu_\Lambda$ it could even lead to a positive NLO
$P$-wave contribution, but that scale may be too small, and the perturbative
calculation will be unreliable.  A physical quantity should be independent of the subtraction
scheme and factorization scale if we perform all order calculation. But at finite order,
the factorization scale dependence does not exactly cancel, leading to scale
ambiguities. The dependence of the decay widths on the factorization scale
$\mu_\Lambda$ induces important theoretical uncertainty. To estimate the theoretical uncertainties caused by the
factorization scale, in Fig. 4, we present the $\mu_\Lambda$ dependence of the $\chi_{c0}$ NLO
total decay width, the $^3S_1^{(8)}$ and $^3P_0^{(1)}$ channel decay widths. From Fig. 4, we can see that
with the decrement of the value of $\mu_\Lambda$ the $^3P_0^{(1)}$-channel
state contribution increases slowly, and this increment of the $^3P_0^{(1)}$-state contribution
is partly compensated by the decrement of $^3S_1^{(8)}$ contribution.
When the scale $\mu_\Lambda$ runs from $0.75~GeV$ to $3.0~GeV$, the $\chi_{c0}$ NLO
total decay width slightly increases with the rise
of $\mu_\Lambda$.

\begin{figure}[!htb]
\begin{center}
\begin{tabular}{cc}
{\includegraphics[width=0.6\textwidth]{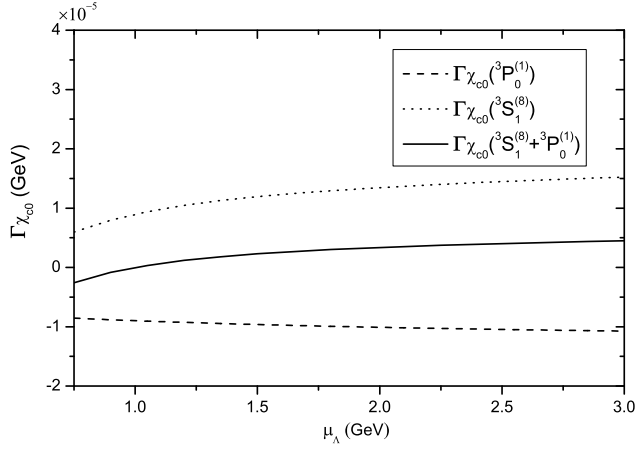}}
\end{tabular}
\end{center}
 \vspace*{-0.8cm}\caption{The $\mu_\Lambda$ dependence of the $\chi_{c0}$ NLO
total decay width, the $^3S_1^{(8)}$ - and $^3P_0^{(1)}$ -channel decay widths.} \label{fl}
\end{figure}

\begin{figure}[!htb]
\begin{center}
\begin{tabular}{cc}
{\includegraphics[width=0.6\textwidth]{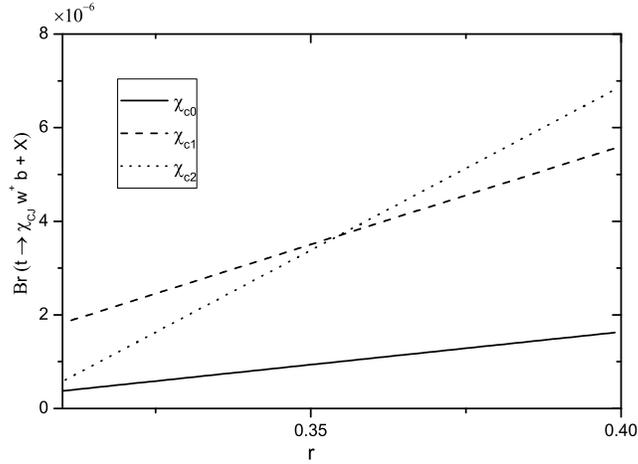}}
\end{tabular}
\end{center}
 \vspace*{-0.7cm}\caption{Branching ratio of the
decay processes $t\to \chi_{cJ}+ W^+ +b$ at NLO as a function of the matrix element ratio $r$.} \label{f4}
\end{figure}

\par
Top-quark decays within the Standard Model are dominated by the
mode $t \to b + W^+$ for $V_{tb}=1$. To get the branching ratio of the
decay processes $t\to \chi_{cJ}+ W^+ +b$, in our work, the Born
approximation decay widths of processes  $t \to b + W^+$ has been used\cite{litop}.
As shown in Fig. 3, the bound of $r>0.301$ is needed to gain a positive
decay width for the process $t\to \chi_{cJ}+ W^+ +b$. In Fig. 5, we present
branching ratios of the decay processes $t\to \chi_{cJ}+ W^+ +b$ at the NLO as a
function of the matrix element ratio $r$ varying from 0.31 to 0.40.
In these processes, the CO contributions are dominant at the LO. When we use the $\overline{MS}$ subtraction
scheme in renormalizing the NRQCD LDMEs $<{\cal
O}^{\chi_{cJ}}[^3S_1^{(8)}]>$ and set factorization scale $\mu_\Lambda=m_c$, the NLO CO give
a big positive correction, and the $^3P_J^{(1)}$ Fock state gives a negative contribution. From our calculation,
we can see that if the value of $r$ favors the Tevatron data and CMS data,
there will not be too much parameter space to accommodate it, the branching ratios of the
decay processes $t\to \chi_{cJ}+ W^+ +b$ are very small, and detecting these processes
will not be easy. Meanwhile, the indirect prompt $J/\psi$ production for the $\chi_{cJ}$
decay will be very small and negligible. If the value of $r$ favors the LHCb data,
detecting $\chi_{c0}$ production from top-quark decay is still very difficult,
but $\chi_{c1}$ and $\chi_{c2}$ production may have the potential to be detected at
the LHC, and the indirect prompt $J/\psi$ production for the $\chi_{cJ}$ may not be so
important as estimated by LO calculation.

\begin{figure}[!htb]
\begin{center}
\begin{tabular}{cc}
{\includegraphics[width=0.6\textwidth]{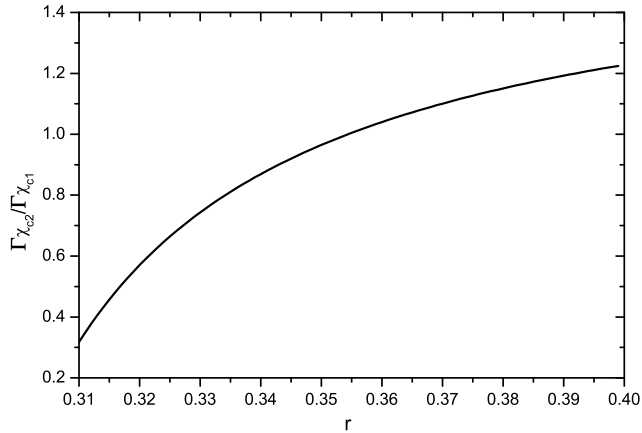}}
\end{tabular}
\end{center}
 \vspace*{-0.7cm}\caption{The ratio $R_{\chi_{c}}=\frac{\Gamma_{\chi_{c2}}}{\Gamma_{\chi_{c1}}}$ at the NLO
 as a function of the the matrix element ratio $r$.} \label{f5}
\end{figure}

In Fig. 6, we also give the ratio $R_{\chi_{c}}=\frac{\Gamma_{\chi_{c2}}}{\Gamma_{\chi_{c1}}}$ at the
NLO as a function of the matrix element ratio $r$ and varying it from 0.31 to 0.40. If we only
consider the LO contribution, we can get the ratio to be $5/3$ by spin counting easily. However, at the NLO,
the ratio $R_{\chi_{c}}$ has been changed significantly for the $^3P_J^{(1)}$-state contribution.
When varying $r$  from 0.31 to 0.40, the ratio $R_{\chi_{c}}$ can be changed from 0.31 to 1.22. With
the growth of the value of $r$, the $^3P_J^{(1)}$-state contribution will become less important,
and the ratio $R_{\chi_{c}}$ will tend to be the spin counting result $5/3$.

\section{Summary}
In conclusion, we have considered the NLO QCD corrections for the
excited charmonium production processes $t\to \chi_{cJ}+ W^+ +b$ in
top-quark decay. These processes are an interesting platform for studying
the heavy quarkonium production mechanism.
We adopt the dimensional regularization to deal
with the UV and IR singularities in our calculation. The Coulomb singularities and
soft singularities in $P$ state are isolated and absorbed into the
NRQCD NLO-corrected operator $<{\cal O}^{\chi_{cJ}}[^3S_1^{(8)}]>$.
After adding all contributing components together, we obtain the
results with UV, IR, and Coulomb safety.
In these processes, the CO contributions are dominant at the LO. When we use the $\overline{MS}$ subtraction
scheme in renormalizing the NRQCD LDMEs $<{\cal
O}^{\chi_{cJ}}[^3S_1^{(8)}]>$ and set factorization scale $\mu_\Lambda=m_c$, the NLO CO gives a
big positive correction, and the $^3P_J^{(1)}$ Fock state gives a negative contribution.
From our calculation, detecting $\chi_{c0}$ production from top-quark decay may
be very difficult, but $\chi_{c1}$ and $\chi_{c2}$ production may have the
potential to be detected at the LHC. Detailed study of the prompt $\chi_{cJ}$ production
from top-quark decay at the LHC is not only  very useful for investigating $J/\psi$
production from top-quark decay but is also important in understanding the heavy
quarkonium production mechanism.

\section{Acknowledgments}
This work was supported in part by the National Natural Science Foundation of China (Grants No.11305001, No.11205003, No.11535002, No.11375171, and No.11275190).

%-------------------------------------------------------------------------------------------------------


\begin{thebibliography}{99}
\bibitem{bbl} G.T. Bodwin, E. Braaten and G.P. Lepage, Phys.
Rev. {\bf D 51} (1995) 1125, erratum ibid. {\bf D 55} (1997) 5853.

\bibitem{longma}
  Y.-Q.~Ma, K.~Wang, and K.-T.~Chao,
  %``$J/\psi (\psi^\prime)$ production at the Tevatron and LHC at ${\cal O}(\alpha_s^4v^4)$ in nonrelativistic QCD,''
  Phys.\ Rev.\ Lett.\  {\bf 106}, 042002 (2011)
  [arXiv:1009.3655 [hep-ph]].
  %%CITATION = ARXIV:1009.3655;%% 19.9.2010

\bibitem{longbu}
  M.~Butensch\"on and B.~A.~Kniehl,
  %``Reconciling $J/\psi$ production at HERA, RHIC, Tevatron, and LHC with NRQCD factorization at next-to-leading order,''
  Phys.\ Rev.\ Lett.\  {\bf 106}, 022003 (2011)
  [arXiv:1009.5662 [hep-ph]].
  %%CITATION = ARXIV:1009.5662;%% 28.09.2010

\bibitem{longgong}
  B.~Gong, L.-P.~Wan, J.-X.~Wang, and H.-F.~Zhang,
  %``Polarization for Prompt J/ψ and ψ(2s) Production at the Tevatron and LHC,''
  Phys.\ Rev.\ Lett.\  {\bf 110}, 042002 (2013)
  [arXiv:1205.6682 [hep-ph]].
  %%CITATION = ARXIV:1205.6682;%%

%\cite{Butenschoen:2009zy}
\bibitem{Butenschoen:2009zy}
  M.~Butensch\"on and B.~A.~Kniehl,
  %``Complete next-to-leading-order corrections to J/psi photoproduction in nonrelativistic quantum chromodynamics,''
  Phys.\ Rev.\ Lett.\  {\bf 104}, 072001 (2010)
  [arXiv:0909.2798 [hep-ph]].
  %%CITATION = ARXIV:0909.2798;%%

%\cite{Butenschoen:2011ks}
\bibitem{Butenschoen:2011ks}
  M.~Butenschoen and B.~A.~Kniehl,
  %``Probing nonrelativistic QCD factorization in polarized $J/\psi$ photoproduction at next-to-leading order,''
  Phys.\ Rev.\ Lett.\  {\bf 107}, 232001 (2011)
  [arXiv:1109.1476 [hep-ph]].
  %%CITATION = ARXIV:1109.1476;%%

%\cite{Ma:2010vd}
\bibitem{Ma:2010vd}
  Y.-Q.~Ma, K.~Wang, and K.-T.~Chao,
  %``QCD radiative corrections to $\chi_{cJ}$ production at hadron colliders,''
  Phys.\ Rev.\ D {\bf 83}, 111503(R) (2011)
  [arXiv:1002.3987 [hep-ph]].
  %%CITATION = ARXIV:1002.3987;%%


%\cite{Butenschoen:2012px}
\bibitem{Butenschoen:2012px}
  M.~Butenschoen and B.~A.~Kniehl,
  %``J/psi polarization at Tevatron and LHC: Nonrelativistic-QCD factorization at the crossroads,''
  Phys.\ Rev.\ Lett.\  {\bf 108}, 172002 (2012)
  [arXiv:1201.1872 [hep-ph]].
  %%CITATION = ARXIV:1201.1872;%%

%\cite{Chao:2012iv}
\bibitem{Chao:2012iv}
  K.-T.~Chao, Y.-Q.~Ma, H.-S.~Shao, K.~Wang, and Y.-J.~Zhang,
  %``$J/\psi$ Polarization at Hadron Colliders in Nonrelativistic QCD,''
  Phys.\ Rev.\ Lett.\  {\bf 108}, 242004 (2012)
  [arXiv:1201.2675 [hep-ph]].
  %%CITATION = ARXIV:1201.2675;%%



%\cite{Shao:2014fca}
\bibitem{Shao:2014fca}
  H.-S.~Shao, Y.-Q.~Ma, K.~Wang, and K.-T.~Chao,
  %``Polarizations of $\chi_{c1}$ and $\chi_{c2}$ in prompt production at the LHC,''
  Phys.\ Rev.\ Lett.\  {\bf 112}, 182003 (2014)
  [arXiv:1402.2913 [hep-ph]].
  %%CITATION = ARXIV:1402.2913;%%

%\cite{Shao:2014yta}
\bibitem{Shao:2014yta}
  H.-S.~Shao, H.~Han, Y.-Q.~Ma, C.~Meng, Y.-J.~Zhang, and K.-T.~Chao,
  %``Yields and polarizations of prompt $\jpsi$ and $\psits$ production in hadronic collisions,''
  J. High Energy Phys.\ 05 ({\bf 2015}) 103
  [arXiv:1411.3300 [hep-ph]].
  %%CITATION = ARXIV:1411.3300;%%

%\cite{Klasen:2004tz}
\bibitem{Klasen:2004tz}
  M.~Klasen, B.~A.~Kniehl, L.~N.~Mihaila, and M.~Steinhauser,
  %``$J/\psi$ plus jet associated production in two-photon collisions at next-to-leading order,''
  Nucl.\ Phys.\  {\bf B713}, 487 (2005)
  [hep-ph/0407014].
  %%CITATION = HEP-PH/0407014;%%

%\cite{Klasen:2004az}
\bibitem{Klasen:2004az}
  M.~Klasen, B.~A.~Kniehl, L.~N.~Mihaila, and M.~Steinhauser,
  %``$J/\psi$ plus prompt-photon associated production in two-photon collisions at next-to-leading order,''
  Phys.\ Rev.\ D {\bf 71}, 014016 (2005)
  [hep-ph/0408280].
  %%CITATION = HEP-PH/0408280;%%

%\cite{Butenschoen:2011yh}
\bibitem{Butenschoen:2011yh}
  M.~Butenschoen and B.~A.~Kniehl,
  %``World data of J/psi production consolidate NRQCD factorization at NLO,''
  Phys.\ Rev.\ D {\bf 84}, 051501(R) (2011)
  [arXiv:1105.0820 [hep-ph]].
  %%CITATION = ARXIV:1105.0820;%%



\bibitem{d1}Zhi-Guo He, Ying Fan and Kuang-Ta Chao, Phys. Rev.
Lett. {\bf 101}, 112001 (2008) [arXiv: 0802.1849].

\bibitem{d2}Ying Fan, Zhi-Guo He, Yan-Qing Ma and Kuang-Ta Chao,
Phys. Rev. {\bf D 80}, 014001 (2009) [arXiv:0903.4572].

\bibitem{comtev} E. Braaten and S. Fleming,
%``Color-Octet Fragmentation and the $\psi'$ Surplus at the Fermilab Tevatron'',
Phys. Rev. Lett. {\bf 74}, 3327 (1995) [hep-ph/9411365].


\bibitem{pnloz}Kuang-Ta Chao, Yan-Qing Ma, Hua-Sheng Shao, Kai Wang, and
Yu-Jie Zhang, Phys. Rev. Lett. {\bf 106}, 042002 (2011) [arXiv:1201.2675] .

\bibitem{polar1}
B.~Gong, L.-P. Wan, J.-X. Wang, and H.-F. Zhang,
\newblock (2013), arXiv:1305.0748;
%%CITATION = ARXIV:1305.0748;%%
CDF, A.~A. Affolder {\em et~al.},
\newblock Phys. Rev. Lett. {\bf 85}, 2886 (2000), arXiv:hep-ex/0004027;
%%CITATION = HEP-EX/0004027;%%
CMS Collaboration, S.~Chatrchyan {\em et~al.},
\newblock Phys.Rev.Lett. {\bf 110}, 081802 (2013), arXiv:1209.2922;
%%CITATION = ARXIV:1209.2922;%%
\newblock (2014), arXiv:1410.8537.
%%CITATION = ARXIV:1410.8537;%%
%\cite{He:2015gla}
\bibitem{heRE}
  Z.~G.~He and B.~A.~Kniehl,
  %``Relativistic corrections to $J/\psi$ polarization in photo- and hadroproduction,''
  Phys.\ Rev.\ D {\bf 92}, 014009 (2015)
  %doi:10.1103/PhysRevD.92.014009
  [arXiv:1507.03883 [hep-ph]].




%\cite{Langenfeld:2009tc}
\bibitem{Langenfeld:2009tc}
  U.~Langenfeld, S.~Moch and P.~Uwer,
  %``New results for t bar t production at hadron colliders,''
  arXiv:0907.2527 [hep-ph].
  %%CITATION = ARXIV:0907.2527;%%

%\cite{Kidonakis:2008mu}
\bibitem{Kidonakis:2008mu}
  N.~Kidonakis and R.~Vogt,
  %``The Theoretical top-quark cross section at the Tevatron and the LHC,''
  Phys.\ Rev.\  D {\bf 78}, 074005 (2008)
  [arXiv:0805.3844 [hep-ph]].
  %%CITATION = PHRVA,D78,074005;%%


%\cite{Liao:2013ika}
\bibitem{Liao:2013ika}
  Q.~L.~Liao, X.~G.~Wu, J.~Jiang, G.~Chen and Z.~Y.~Fang,
  %``Heavy quarkonium production through the top-quark decays via flavor changing neutral currents,''
  arXiv:1304.1303 [hep-ph].
  %%CITATION = ARXIV:1304.1303;%%

%\cite{Qiao:1996rd}
\bibitem{Qiao:1996rd}
  C.~F.~Qiao, C.~S.~Li and K.~T.~Chao,
  %``Top-quark decays into heavy quark mesons,''
  Phys.\ Rev.\ D {\bf 54}, 5606 (1996)
  %doi:10.1103/PhysRevD.54.5606
  [hep-ph/9603275].
  %%CITATION = doi:10.1103/PhysRevD.54.5606;%%

\bibitem{qiao} Cong-feng Qiao, Kuang-Ta Chao,
%"Color-Octet Charmonium Production in top-quark Decays",
Phy. Rev. {\bf D 55}, 2837 (1997) [arXiv:9606462].

%\cite{Yuan:1997yh}
\bibitem{Yuan:1997yh}
  F.~Yuan, C.~F.~Qiao and K.~T.~Chao,
  %``Color singlet and color octet $J/\psi$ production in top-quark rare decays,''
  Phys.\ Rev.\ D {\bf 57}, 610 (1998)
  %doi:10.1103/PhysRevD.57.610
  [hep-ph/9709400].
  %%CITATION = doi:10.1103/PhysRevD.57.610;%%

%\cite{Sun:2010rw}
\bibitem{Sun:2010rw}
  P.~Sun, L.~P.~Sun and C.~F.~Qiao,
  %``The Next-to-Leading Order Corrections to top-quark Decays to Heavy Quarkonia,''
  Phys.\ Rev.\ D {\bf 81}, 114035 (2010)
  %doi:10.1103/PhysRevD.81.114035
  [arXiv:1003.5360 [hep-ph]].
  %%CITATION = doi:10.1103/PhysRevD.81.114035;%%

\bibitem{litop}
Song Mao, Li Gang, Zhou Ya-Jin, Guo Jian-You, and Ma Zheng-Wei, Phys. Rev. D {\bf 91}, 116004 (2015).

\bibitem{p2}
  A.~Petrelli, M.~Cacciari, M.~Greco, F.~Maltoni and M.~L.~Mangano,
  %``NLO production and decay of quarkonium,''
  Nucl.\ Phys.\ B {\bf 514}, 245 (1998)
  %doi:10.1016/S0550-3213(97)00801-8
  [hep-ph/9707223].
  %%CITATION = doi:10.1016/S0550-3213(97)00801-8;%%
  %207 citations counted in INSPIRE as of 31 Oct 2016



\bibitem{coulomb} M. Kramer,
%``QCD corrections to inelastic $J/\psi$ photoproduction'',
 Nucl. Phys. {\bf B 459}, 3 (1996).




\bibitem{pss}
  W. T. Giele and E. W. N. Glover, Phys. Rev. {\bf D46} (1992) 1980;
  W. T. Giele, E. W. Glover and D. A. Kosower, Nucl. Phys. {\bf B403} (1993) 633;
  S. Keller and E. Laenen, Phys. Rev. {\bf D59} (1999) 114004.

\bibitem{Eichten:1995PRD}
E.J. Eichten and C. Quigg, Phys. Rev. {\bf D52}, 1726 (1995).

\end{thebibliography}
\end{document}